\documentclass[superscriptaddress,pre,twocolumn,amsmath,amssymb,floatfix]{revtex4}
\usepackage{graphicx}% Include figure files
\usepackage{dcolumn}% Align table columns on decimal point
\usepackage{bm}% bold math

\begin{document}

%\preprint{APS/123-QED}

\title{Collective Helping and Bystander Effects in Coevolving Helping Networks}
% Force line breaks with \\

\author{Hang-Hyun Jo} %\email{h2jo@kias.re.kr}
\affiliation{School of Physics, Korea Institute for Advanced
Study, Seoul 130-722, Korea}

\author{Hyun Keun Lee} %\email{hkleepp@gmail.com}
\affiliation{School of Physics, Korea Institute for Advanced
Study, Seoul 130-722, Korea} \affiliation{Department
of Physics, BK21 Physics Research Division, Sungkyunkwan
University, Suwon 440-746, Korea}

\author{Hyunggyu Park} %\email{hgpark@kias.re.kr}
\affiliation{School of Physics, Korea Institute for Advanced
Study, Seoul 130-722, Korea}

\date{\today}% It is always \today, today,
             %  but any date may be explicitly specified

\begin{abstract}
We study collective helping behavior and bystander effects
in a coevolving helping network model. A node and a link of the
network represents an agent who renders or receives help and a
friendly relation between agents, respectively. A helping trial of
an agent depends on relations with other involved agents and its
result (success or failure) updates the relation between the
helper and the recipient. We study the network link dynamics and
its steady states analytically and numerically. The full phase
diagram is presented with various kinds of active and inactive
phases and the nature of phase transitions are explored. We find
various interesting bystander effects, consistent with the field
study results, of which the underlying mechanism is proposed.

%We propose the underlying mechanism for
% phase transitions of link density in the coevolving
%helping network model. Each node and each link of network
%represent an agent who renders or receives help and the
%expectation of help between two agents, respectively. The helping
%behavior of agent depends on his or her relations with the other
%involved agents, and leads to the evolution of links according to
%the result of help. The active and the inactive phases and the
%transitions among them are identified by mean field analysis and
%confirmed by numerical simulations. We also observe the various
%collective bystander effects.

\end{abstract}

% PACS, the Physics and Astronomy
                             % Classification Scheme.
% 87.23.Ge - Dynamics of social systems
% 89.65.-s - Social and economic systems
% 89.90.+n - Other topics in areas of applied and interdisciplinary physics
% 68.35.Rh - Phase transitions and critical phenomena

%Use showkeys class option if keyword
                              %display desired
\maketitle

\section{Introduction}\label{sect:intro}

In recent years, statistical mechanics has played an important
role in quantitative investigation of the social, economic, and
psychological phenomena~\cite{CMV}. Some social phenomena bear the
resemblance to the physical processes in a sense that the
macroscopic complex patterns emerge from the interaction of a
large number of microscopic constituents. Along this line, one of
the interesting subjects to physicists is opinion dynamics, most
of which are concerned either with agent based model
studies~\cite{Stauffer2005,Sznajd2000} or with the analysis of
real data such as election
results~\cite{CostaFilho1999,Fortunato2007}. For the model studies
the effect of interaction structure has been considered in terms
of the complex network theory~\cite{ANB}. More recently, various
coevolving networks with the agents interacting on those networks
have been proposed and studied~\cite{GFSHPB}. The coevolution of
networks with agents may be justified for a broad range of social
phenomena, because each agent may construct or reconstruct his or
her own neighborhoods in response to the interaction with other
agents.

In this paper, we study a coevolving helping network model, known
as the rescue model~\cite{Jo2006}. This model was motivated by the
{\em bystander effect} observed in social psychology: The
witnesses are less likely to intervene in the emergency situation
when there are more witnesses. The bystander effect was introduced
by Latan\'e and his colleagues in the late
1960's~\cite{Latane1969}. Since then, the prosocial and helping
behavior was investigated extensively from micro- to macro-level
perspectives~\cite{Penner2005}. The meso-level perspective
corresponds to the study of helper-recipient dyads. We would like
to see how the collective helping behavior at the level of society
emerges from the repeated helper-recipient interactions. We adopt
the coevolutionary dynamics consisting of two stages such as (a)
an agent's behavior depending on the relations with other involved
agents and (b) the relation update resulting from the agent's
behavior.

The assumptions of the model are based on the results of social
psychological experiments in the
laboratory~\cite{Latane1969,Rosen1987}: First, the witnesses who
had an acquaintance with the victim  were faster to report the
victim's distress than did the other witnesses. Second, the
witnesses who had friends among themselves responded to the
emergency situation faster. Thus the relations among the involved
agents definitely affected their helping behaviors. For
simplicity, we assume that the various relations among agents,
such as friendship or short acquaintances, can be modeled by a
simple (unweighted and undirected) link between nodes of a
network, where a node represents an agent.

We interpret the link as one node's expectation of help by the
other node. The link density of a network can be interpreted as
the aggregate expectation of help in a society. A new link between
the intervening witness (helper) and the victim (recipient) may be
created as a result of the successful intervention. Or the
existing link between them may be severed when the intervention
fails, because the failure may reduce the future expectation of
help between each other. Furthermore, to build the feasible model
we adopt the cost-reward model for the witness's arousal to the
emergency~\cite{Piliavin1982}. One of the basic assumptions is
that the degree of arousal that a witness perceives is a
monotonically increasing function of the perceived severity and
clarity of the emergency. Here we will use the more explicit term,
i.e. the degree of willingness to intervene, rather than the
degree of arousal.

In the previous works on the rescue model~\cite{Jo2006}, the
authors investigated the effect of the number of involved
witnesses $k$ on the collective helping behavior. They reported a
{\em non-monotonic} variation of the link density (aggregate expectation
of help) with $k$ at some moderate values of the model parameters,
which are partly consistent with the field study
results~\cite{Amato1983}. In this work, we present the full phase
diagram through the comprehensive analytic study of the model with
small $k$ in the whole parameter space.  We identify the various
kinds of active and inactive phases and analyze the nature of
phase transitions. Furthermore, we study the fluctuations which
are responsible for  finite-size effects and  correlations.
Numerical simulation results for large $k$ reveal various kinds
({\em normal, reverse, complex}) of
bystander effects depending on the parameter values. We propose
the underlying mechanism for these bystander effects,
which is consistent with the numerical and analytical results.

%The other model
%parameters were fixed as the moderate numbers.
% Since they already
%considered in the model the adverse effect among the witnesses who
%are strangers, their results should not be too surprising. To
%clarify the implication of the previous works we distinguish the
%adverse effect at the dyadic level from the collective and
%resultant bystander effect at the social level. The link between
%those two levels is not straightforward because of the complicated
%interplay among the model parameters. Precisely, the aggregate
%expectation of help (link density) turned out to behave
%non-monotonically with the number of witnesses. In this paper, we
%study the same model but more comprehensively by scanning the
%whole parameter space. We identify the active and the inactive
%phases and analyze the phase transitions among them when the
%number of involved witnesses is given as small. Then we can obtain
%the insight of the case with the larger numbers of witnesses.
%Precisely, we investigate the collective bystander effects by
%observing how the aggregate expectation of help depends on the
%number of witnesses per accident.

This paper is organized as following. In Sec.~\ref{sect:model},
the coevolving helping network model is briefly introduced. In
Sec.~\ref{sect:pt}, the active and inactive phases and the
transitions among them are identified by the analytic calculations
and confirmed by numerical simulations. In
Sec.~\ref{sect:bystander}, the various bystander effects are
discussed. Finally, we summarize the results in
Sec.~\ref{sect:con}.

\section{\label{sect:model}Model}

We briefly introduce the coevolving helping network (CHN)
model~\cite{Jo2006}. In the CHN, a node represents an agent and a
link between two nodes represents a {\em friendly} relation
between two agents. The links are unweighted and undirected, so
the network is defined by the symmetric adjacency matrix
$\{\rho_{ij}\}$ where $\rho_{ij}=1$ if two nodes $i$ and $j$ are
connected by a link and $0$ otherwise.

At each time step, an accident (emergency) occurs involving a
randomly chosen agent $v$ (victim) and also randomly chosen $k$
agents (witnesses) from a population of $N$ agents (nodes). The
set of $k$ agents for the victim $v$ is denoted by $\Lambda_v$.
Each accident carries the degree of its severity represented by
$q$, which is randomly drawn uniformly from $[0,1)$. Each witness
$i\in \Lambda_v$ needs to overcome, at least, its nonnegative
potential cost $c_i$ to intervene in an accident (rescue attempt).
Furthermore, the possibility of the rescue attempt by the witness
should be enhanced by the friendly relation between the victim and
the witness as well as the number of friends of the witness in the
other $k-1$ witnesses.

Summing up all together, we may write the degree of willingness
$x_{iv}$ of the witness $i$ to intervene in the accident occurred
at time $t$ on the victim $v$ as
\begin{equation}
%x_{vi}(t)=q+a \rho_{vi}(t) + b \sum_{j\in N_v \setminus
%i}[2\rho_{ij}(t)-1] -c_i.\label{eq:xvi}
x_{iv}(t)=q+a \rho_{iv}(t) + b \sum_{j\in \Lambda_v^i}
[2\rho_{ij}(t)-1] -c_i ,\label{eq:xvi}
\end{equation}
where $a\geq 0$, $b\geq 0$, and $\Lambda_v^i$ denotes the set
$\Lambda_v$ excluding the witness $i$. The interaction term
between witnesses enhances $x_{iv}$ if the witness $i$ has more
friends than strangers in $\Lambda_v^i$, and vice versa. The
magnitude of the interaction term increases with $k$ as it can
vary from $b(k-1)$ (all other witnesses are friends of the witness
$i$) to $-b(k-1)$ (all other witnesses are strangers to the
witness $i$). So we expect that the interaction term may dominate
the evolving dynamics over all other terms for sufficiently large
$k$.

We select the witness $i\in \Lambda_v$ with the largest value of
$x_{iv}$ as the intervener~\cite{Jo2006}. If there are more than
one witnesses with the largest $x_{iv}$, one of them is selected
randomly. The intervention attempt would succeed if $x_{iv}\ge 0$
and fail otherwise, which is followed by the link (relation)
update between the victim and the intervener. If successful, the
friendly relation will be set up or maintained between the victim
and the intervener. Otherwise, they become or remain strangers to
each other. The link update dynamics is summarized as
\begin{equation}
\rho_{iv}(t+\Delta t)=\theta \left[x_{iv}(t)\right], \label{eq:evol}
\end{equation}
where $\theta(x)$ is a Heaviside step function with $\theta(x)=1$
for $x\ge 0$ and $0$ for $x<0$. We choose the time step $\Delta
t=1/L$ for convenience where $L=\frac{N(N-1)}{2}$ is the total
number of possible links between $N$ nodes. At the next time step,
the above procedure is repeated and the time $t$ is incremented by
1 after $L$ such iterations.

We focus on the link density $\rho(t)$ and the hole density $u(t)$
as the order parameters, respectively:
\begin{equation}
\rho(t)\equiv1-u(t)\equiv\frac{1}{L}\sum_{i<j}{\rho_{ij}(t)},\label{eq:rho}
\end{equation}
which should be proportional to the aggregate mean expectation of
help (intervention) in a society, which can be one of the
important social features.

Finally, in this work, we assume $c_i=c\ge 0$ for all $i$ for
simplicity, which may be valid for a sufficiently homogeneous
population.

\section{\label{sect:pt}Nonequilibrium Phase Transitions}

\subsection{The case with $a=0$ and $b=0$}

%First, consider the case with $b=0$.
The $a=b=0$ case is special. Every witness's degree of willingness
is identical as $x_{iv}=q-c$ and the intervener is randomly
selected among witnesses, regardless of the number of witnesses
$k$. The probability for the successful intervention ($x_{iv}\ge
0$) should be equal to $1-c$ for $0\le c\le 1$. Thus,
probabilistically, the number of links increases by 1 with the
rate $W_+=(1-c)(1-\rho)$ and decreases by 1 with the rate
$W_-=c\rho$. For $c>1$, $W_+=0$ and $W_-=\rho$.

In the mean-field (MF) framework ignoring the stochastic temporal
noise, the rate equation for the link density in the
time-continuum limit is given as
\begin{equation}\label{rate_0}
\frac{d\rho}{dt}=W_+-W_- ,
\end{equation}
where
\begin{eqnarray}\label{tran_rate_0}
W_+&=&F(1-c)(1-\rho), \nonumber\\
W_-&=&F(c)\rho,
\end{eqnarray}
with  $F(x)=x$ for $0<x<1$, $1$ for $x\ge 1$,
and $0$ for $x\le 0$. It is easy to show that $\rho(t)$ behaves as
\begin{equation}
\rho(t)=\rho(0)e^{-t} + \rho_\infty (1-e^{-t}),\label{eq:sol}
\end{equation}
with the steady-state density
$\rho_\infty\equiv\lim_{t\rightarrow\infty}\rho(t)$ as
\begin{equation}
\rho_\infty=F(1-c),
\end{equation}
which implies that there is a nonequilibrium phase transition at
$c=1$ from an active phase ($\rho_\infty=1-c$ for $c<1$) into an
inactive phase ($\rho_\infty=0$ for $c\ge 1$). The order parameter
exponent $\beta$ is defined near the transition point in the
active side as
\begin{equation}
\rho_\infty\simeq \varepsilon^\beta,
\end{equation}
with the reduced coupling constant $\varepsilon$ which measures
the distance from the transition point. In this case,
$\varepsilon=1-c$ and $\beta=1$.

In the inactive phase, the steady state is the vacuum (no link
state) where the dynamics becomes completely dead, which is called
as an {\em absorbing} (trapped) state. However, this transition is
different from other well-known absorbing phase transitions
\cite{abtr}, since the vacuum is not absorbing in the active
phase. Furthermore, the system at the transition point ($c=1$) is
not critical, but absorbing: The link density decays exponentially
($\rho(t)=\rho(0)e^{-t}$).

One may derive the exact Langevin equation for large $L$,
including the stochastic noise (in the Ito sense \cite{Ito})
through the usual Fokker-Planck formulation as
\begin{equation}\label{exact1}
\frac{d\rho}{dt}=f(\rho) + \sqrt{g(\rho)} \xi(t),
\end{equation}
where $f(\rho)=W_+ - W_-$ and $g(\rho)=(W_+ + W_-)/L$. $\xi(t)$ is
a white noise with zero mean satisfying
$\langle\xi(t)\xi(t^\prime)\rangle=\delta(t-t^\prime)$ where
$\langle\cdots\rangle$ is the noise average. The drift function
$f(\rho)=(1-c)-\rho$ and the multiplicative factor
$g(\rho)=[(1-c)+(2c-1)\rho]/L$ for $c<1$, while $f(\rho)=-\rho$
and $g(\rho)=\rho/L$ for $c\ge 1$. In the $L=\infty$ limit, the
noise term becomes negiligible and Eq.~(\ref{rate_0}) becomes
exact.

The fluctuations due to the stochastic noise can be derived, using
the Ito calculus \cite{Ito} such that
\begin{equation}\label{Itoc}
\frac{d \langle G(\rho)\rangle}{dt}=\left\langle f(\rho)
\frac{\partial G}{\partial\rho} \right\rangle + \frac{1}{2}
\left\langle g(\rho) \frac{\partial^2G}{\partial\rho^2}
\right\rangle,
\end{equation}
where $G$ is a general function of $\rho$.
Then, the order parameter fluctuation $\chi$ defined as
\begin{equation}
\chi=L\left[ \langle \rho^2 \rangle -\langle \rho \rangle^2 \right],
\end{equation}
behaves in the steady state as
\begin{equation}
\chi_\infty=\left\{
\begin{array}{cll}
c(1-c) &\mbox{ for } c<1 &\\
0  &\mbox{ for } c\ge 1&
\end{array}
\right. .
\end{equation}
Note that the fluctuation vanishes as the system approaches the
transition point ($c=1$). The fluctuation exponent $\gamma^\prime$
defined as $\chi\sim \varepsilon^{-\gamma^\prime}$ is given as
$\gamma^\prime=-1$.

\subsection{The case with $a>0$ and $b=0$}

For $a>0$ and $b=0$, we have the same rate equation as in
Eq.~(\ref{rate_0}) with the link creation and annihilation rates
as
\begin{eqnarray}\label{tran_rate}
W_+&=&F(1-c)(1-\rho)^k, \nonumber\\
W_-&=&F(c-a)[1-(1-\rho)^k],
\end{eqnarray}
%Each $W$ appears as a product of its prefactor (function of
%control parameters) and its combinatorial part (function of link
%density).
respectively. Note that the $a=0^+$ limit is singular except the case of $k=1$.

\begin{figure}[t]
\centerline{\includegraphics[scale=.8]{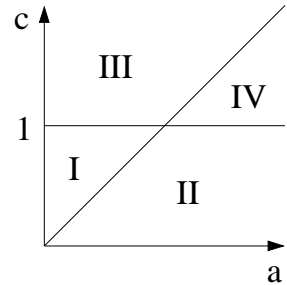}}
\caption{The phase diagram for the case with $b=0$. The parameter
space $(a,c)$ is divided into four regions by two transition lines
of $c=1$ and $c=a$.} \label{fig:phaseb0}
\end{figure}

\begin{figure}[t]
\centerline{\includegraphics[scale=.8]{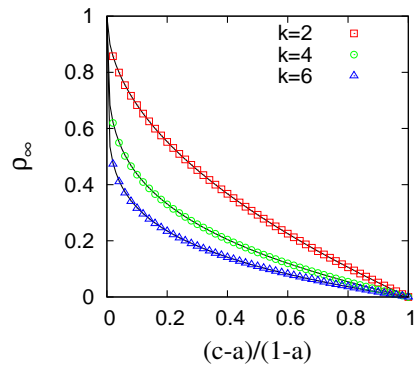}}
\caption{(Color online) The steady-state link density
$\rho_\infty$ along the $a=0.5$ line for $k=2, 4,$ and $6$.}
\label{fig:order}
\end{figure}

It is straightforward to analyze the rate equation, which yields
the steady-state density and its long-time dynamics. The parameter
space  $(a,c)$ is divided into four different regions separated by
two transition lines of $c=1$ and $c=a$ (Fig.~\ref{fig:phaseb0}).
In region I ($a<c<1$), we have a finite link density as
\begin{equation}\label{rho2}
\rho_\infty=1-[(c-a)/(1-a)]^{1/k}
\end{equation}
(active phase). In region II ($c< a$ and $c<1$), we find the fully
connected network; $\rho_\infty=1$ ({\em paradise} phase). In
region III ($c>a$ and $c>1$), the network becomes completely
disconnected; $\rho_\infty=0$ ({\em isolation} phase). In region
IV ($1<c<a$), both transition rates $W_+=W_-=0$ and no dynamics
occurs; $\rho(t)=\rho(0)$. Of course, the region IV is unphysical,
so we focus on the three other regions and the transition lines
between them.

Approaching the I-III transition line ($c=1^-$) inside the active
phase, the order parameter scales as
\begin{equation}
\rho_\infty\simeq A_1\varepsilon_1^{\beta_1},
\end{equation}
with $A_1=1/[k(1-a)]$, $\varepsilon_1=1-c$, and $\beta_1=1$. Near
the I-II transition line ($c=a^+$), the hole density $u=1-\rho$
plays a role of the order parameter, which scales as
\begin{equation}
u_\infty\simeq { A_2}\varepsilon_2^{\beta_2},
\end{equation}
with $A_2=(1-a)^{-1/k}$, $\varepsilon_2=c-a$, and $\beta_2=1/k$.

Note that the steady-state link density $\rho_\infty$ becomes
smaller as $k$ increases. In fact, one can easily show from
Eq.~(\ref{rho2}) that this holds for the whole region of the
active phase, see Fig.~\ref{fig:order}. It implies that the
society becomes less friendly (lower link density) with large $k$
(more crowded society like big cities) in the long-time limit
({\em normal} bystander effect). Therefore, one may regard the
normal bystander effect as the result of iterative intervention
attempts at the level of society.

From the dynamic rules (transition rates) of our model,
Eq.~(\ref{tran_rate}), it is easy to understand how this happens.
A new link may be created only when there are no existing links
between a victim and any of $k$ witnesses, which we call the {\em
$k$-hole constraint}. (Note that this applies only when $a>0$.)
The probability of finding this situation is proportional to
$(1-\rho)^k$, which monotonically decreases with $k$. So one can
expect that the steady-state link density should be smaller with
larger $k$. However, with interactions between witnesses ($b\neq
0$), the $k$-hole constraint is not absolutely necessary to create
a new link, which will be discussed later.

Furthermore, suppose we are allowed to control $c$, for example,
decrease $c$ by giving all agents the same incentive for
intervention. The efficiency of our policy might be measured by
the slope of the link density, i.e. $e\equiv
-\frac{\partial\rho}{\partial c}$. As can be easily seen in
Fig.~\ref{fig:order}, the policy becomes quite effective near the
paradise phase and more effective for larger $k$. However, it will
be less efficient near the isolation phase and much lesser for
larger $k$. These results imply that the incentive policy would
not work well in unfriendly and crowded communities like big and
rapidly developed cities, but may work efficiently in friendly and
crowded communities like a guild.
%small countryside towns.

Now, we study the long-time dynamics in various regions. In the
regions I and III, the link density decays exponentially:
$\rho(t)-\rho_\infty \sim e^{-t/\tau}$ with the characteristic
time $\tau$ such that $\tau^{-1}=k(1-a)^{1/k} (c-a)^{1-1/k}$ in
the active phase and $\tau^{-1}= k(c-a)$ for $c<a+1$ or $k$ for
$c\ge a+1$, respectively in the isolation phase. Along the I-III
transition line, the link density still decays exponentially with
$\tau^{-1}=k(1-a)$, which implies that this transition line is not
critical but absorbing, i.e.~ the $c=1$ line belongs to the
isolation phase. Approaching the I-II transition line ($c=a^+$),
the relaxation time diverges as $\tau\sim
\varepsilon_2^{-\nu_\parallel}$ with the relaxation time exponent
$\nu_\parallel=1-1/k$ for $k>1$. So the incentive policy works
efficiently in this region, but its effect will show up quite
slowly.

In the region II (paradise phase), the long-time decay dynamics
follows a much slower power-law: the hole density $u(t) \sim
t^{-\delta}$ with the decay exponent $\delta=1/(k-1)$ for $k>1$
except that $u(t) \sim e^{-(1-c)t}$ for $k=1$. With larger $k$,
the dynamics becomes extremely slower. The I-II transition line
belongs to the paradise phase, meaning that the system exhibits
the same power-law decay dynamics.

As in Eq.~(\ref{exact1}), one may write the exact Langevin
equation including the stochastic noise with the transition rates
given by Eq.~(\ref{tran_rate}). Near the I-III transition line,
one can expand in powers of $\rho$ and obtain that
$f(\rho)=(1-c)-(1-a)k\rho$ and $g(\rho)=[(1-c)+(2c-a-1)k\rho]/L$
in the active phase, while $f(\rho)=-(c-a)k\rho$ and
$g(\rho)=(c-a)k\rho/L$ $(c<a+1)$ or $k\rho/L$ $(c\ge a+1)$ in the
isolation phase. Then, the steady-state fluctuations are
analytically obtained,  similar to the last subsection, as
$\chi_\infty\simeq B_1\varepsilon_1^{-\gamma_1^\prime}$ with
$B_1=(c-a)/[k(1-a)^2]$ and $\gamma_1^\prime=-1$ in the active side
and $\chi_\infty=0$ in the isolation side.

Near the I-II transition line, we need to write the Langevin
equation in terms of the hole density $u$ as
\begin{equation}\label{exact2}
\frac{du}{dt}={\tilde f}(u) + \sqrt{{\tilde g}(u)} \xi(t),
\end{equation}
where ${\tilde f}(u)=(c-a)-(1-a)u^k$ and ${\tilde
g}(u)=[(c-a)+(1+a-2c)u^k]/L$ in the active phase, and ${\tilde
f}(u)=-(1-c)u^k$ and ${\tilde g}(u)=(1-c)u^k/L$ in the paradise
phase.

One can derive the $k$-th order fluctuation  analytically, using
the Ito calculus of Eq.~(\ref{Itoc}) as
\begin{equation}
\chi_\infty^{(k)}\equiv L\left[ \langle u^{k+1} \rangle -\langle
u^k \rangle \langle u \rangle \right]
=B_2\varepsilon_2^{-\gamma_2^{\prime (k)}},
\end{equation}
where $B_2=(1-c)/(1-a)^2$ and $\gamma_2^{\prime (k)}=-1$. From the
simple scaling theory, one can relate the ordinary mean-square
fluctuation $\chi_\infty=\chi_\infty^{(1)}$ with
$\chi_\infty^{(k)}$ through the exponent relation of
$\gamma_2^\prime=\gamma_2^{\prime (k)} + (k-1)\beta_2$. Hence, we
find $\gamma_2^\prime=-1/k$ with $\beta_2=1/k$.

\begin{figure}[t]
\centerline{\includegraphics[scale=.8]{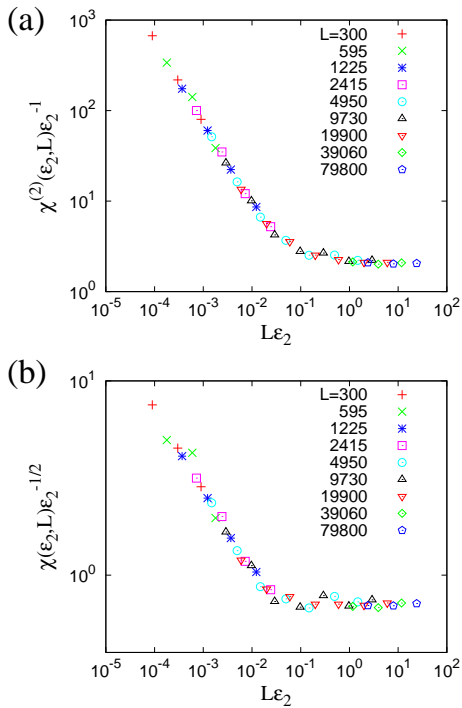}}
\caption{(Color online) The scaling collapse of numerical data for
the fluctuations (a) $\chi_\infty^{(2)}$ and (b) $\chi_\infty$ at
$k=2$ and $a=0.5$ in the active phase ($\varepsilon_2>0$).}
\label{fig:fss}
\end{figure}

Finally, we discuss the finite-size effects near the I-II
transition line. The standard finite-size-scaling (FSS) theory
predicts
\begin{equation}
\chi_\infty^{(k)}=\varepsilon_2^{-\gamma_2^{\prime (k)}}
\phi^{(k)}\left(\varepsilon_2 L^{1/\bar\nu_2}\right),
\end{equation}
where the scaling function behaves as $\phi^{(k)}(x)\sim O(1)$ for large
$x$ and $\phi(x)^{(k)}\sim x^{\gamma_2^{\prime (k)}}$ for small $x$.
Fig.~\ref{fig:fss}(a) shows an excellent collapse of various numerical
finite-size data for
$\chi_\infty^{(2)}$ at $k=2$ and $a=1/2$ onto a scaling curve $\phi^{(2)}(x)$,
with the choice of the FSS exponent $\bar\nu_2=1$. The same is true for
the ordinary fluctuation $\chi_\infty$ with the scaling function $\phi (x)$
which behaves as $\phi(x)\sim O(1)$ for large
$x$ and $\phi(x)\sim x^{\gamma_2^{\prime}}$ for small $x$ (Fig.\ref{fig:fss}(b)).
It follows that the order parameter $u$ should scale as $u\sim L^{-\beta_2/\bar\nu_2}=L^{-1/2}$
at $\varepsilon_2=0^+$. Note that the FSS exponent $\bar\nu_2$ does not satisfy
the standard hyperscaling relation as $\gamma_2^\prime+2\beta_2=\bar\nu_2$.

In the paradise phase including the I-II boundary, only the link
creation process is allowed (no link annihilation dynamics). So
there can be infinitely many absorbing states where each node has
some unconnected links (holes) and the number of those holes is
less than $k$ for all nodes. Then, as there are no configurations
satisfying the $k$-hole constraint at all nodes, a link creation
attempt is impossible and the system is frozen dynamically. The
hole density in such absorbing states should be proportional to
$kN/L$ or equivalently $u\sim L^{-1/2}$ in this region
($\varepsilon_2\le 0$), which is also confirmed numerically.
In addition, there is a discontinuous jump
in the hole density for finite systems across the I-II transition line,
due to the presence of this frozen dynamics in the paradise phase.

\subsection{The case with $b>0$}

\begin{figure}
\centerline{\includegraphics[scale=.8]{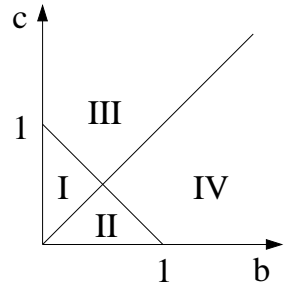}}
\caption{The phase diagram for the case with $a=0$ and $k=2$.}
\label{fig:pd_b0}
\end{figure}

We now consider more realistic cases with the interaction term
($b>0$). In this case, each witness's degree of willingness in
Eq.~(\ref{eq:xvi}) is enhanced if he has more friends than
strangers among the other witnesses and vice versa. Then, one may
naively expect that the interaction drives a high link-density
network   to become a higher one, and a low link-density network
to become a lower one. As a result, both the isolation phase and
the paradise phase would expand into the active phase. This is
true at $a=0$ and, for sufficiently large $b$ $(\ge 1/[2(k-1)])$,
the active phase squeezes down to disappear (see, for example,
Fig.~\ref{fig:pd_b0}).

However, for $a>0$, the $k$-hole constraint is in effect, which
suppresses the link density increase in general. Its effect is
particularly big in the high density networks, so the paradise
phase may shrink. At the same time, the interaction term may
loosen the constraint for large $k$ and thus the naive picture can
be restored. In fact, the $k$-hole constraint becomes loose for $a
< 2b (k-2)$, when one of the witnesses without a link to the
victim may take a higher degree of willingness than one with a
link to the victim. So, a link creation dynamics is possible for a
configuration not satisfying the $k$-hole constraint.

At $k=2$, the $k$-hole constraint remains intact for any value of
$b$ and $a>0$. The link density decreases even in the high
link-density networks. Then, the active phase expands into the
paradise phase and also extends to the large $a$ region. In
contrast, for $k\ge 3$ and small $a$ $(<2b(k-2))$, we find that
the naive expectation holds and the active phase shrinks due to
the expansion of both the isolation and the paradise phase.

As discussed in the last subsection for $b=0$, the $k$-hole
constraint is the key factor for the normal bystander effect:
the steady-state link density $\rho_\infty$ decreases
monotonically with $k$. For $b>0$, however, as $k$ increases, the
constraint becomes looser and the interaction term becomes
stronger $(\sim b(k-1)(2\rho-1))$ in the high density regime
($\rho>1/2$). In this case, we expect the density increase with
$k$ for sufficiently large $k$ after the ordinary density decrease
with $k$ for small $k$. These non-monotonic bystander effects will be
discussed more in the next section.

Now, we consider the $k=2$ and $k\ge 3$ case in more details.
\subsubsection{$k=2$}

First, consider the $a=0$ case. One may easily show that the
transition rates are given as
\begin{eqnarray}\label{tran_rate_k2_a0}
W_+&=&(1-\rho)\left[F(1-c-b)(1-\rho)+F(1-c+b)\rho\right], \nonumber\\
W_-&=&\rho\left[F(c+b)(1-\rho)+F(c-b)\rho\right],
\end{eqnarray}
respectively. We find the phase diagram as in Fig.~\ref{fig:pd_b0}
where the four different phases are separated by two transition
lines of $c=b$ and $c=1-b$. As expected, the active phase shrinks
with increasing $b$ by invasion of both the isolation and the
paradise phase. In the active phase, the steady-state link density
is $\rho_\infty=(1-b-c)/(1-2b)$, which scales linearly
($\beta_1=1$) near the I-III transition line $(c=1-b)$ and the
hole density $u_\infty$ also scales linearly ($\beta_2=1$) near
the I-II transition line $(c=b)$.

With finite $a>0$, the $k$-hole constraint is in effect and the
transition rates are modified as
\begin{eqnarray}\label{tran_rate_k2}
W_+&=&(1-\rho)^2\left[F(1-c-b)(1-\rho)+F(1-c+b)\rho\right], \nonumber\\
W_-&=&\rho(2-\rho)\left[F(c-a+b)(1-\rho)+F(c-a-b)\rho\right], \nonumber\\
\end{eqnarray}
respectively. Note that the $a=0^+$ limit is again singular due to
the $k$-hole constraint.

\begin{figure}[t]
\centerline{\includegraphics[scale=.8]{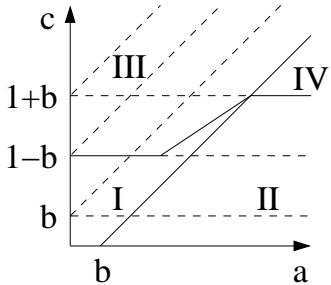}}
\caption{The phase diagram for the case with $k=2$ and
$0<b<\frac{1}{2}$. The parameter space $(a,c)$ is divided by four
transition lines such as $c=1\pm b$, $c=a-b$, and $c=(1+2a-b)/3$.
The dashed lines are 7 non-analytic lines for the transition
rates.} \label{fig:phaseb}
% and the dotted lines are only to guide the eyes.
\end{figure}

The parameter space $(a,c)$ is divided into 14 different regions
by 7 lines such as $c=1\pm b$, $c=b$, $c=a + (1\pm b)$, and
$c=a\pm b$, where the transition rates change abruptly due to the
non-analyticity of the function $F(x)$ at $x=0$ and $1$. The
divided regions are grouped into the four phases as shown in
Fig.~\ref{fig:phaseb}, except that one central region is divided
into two phases of I and III for $0<b<1/2$. As expected, the link
density decreases with $b$ for most regions~\cite{exception},
mainly due to the $k$-hole constraint. The isolation phase III
invades the active phase I and the paradise phase II retreats. We
find the similar result for $b>1/2$ (not shown here).

The active phase I comprises of four sub-phases where the order
parameter $\rho_\infty$ behaves differently:
\begin{equation}
\begin{split}
&1-\rho_\infty=\\
&\left\{
\begin{array}{ll}
\frac{b+\sqrt{b^2+(1-a)(c-a-b)}}{1-a}&\textrm{for $c\geq a+b$ (Ia)}\\
\frac{c-a+b}{1+a-2c}&\textrm{for $c\geq 1-b$ (Ib)} \\
\frac{1-\sqrt{1+4a^2-4ab-4ac}}{2a} &
\textrm{for $c\leq b$ (Ic)}\\
\frac{1+b-c-\sqrt{4a^2-3b^2+2b+1-2(4a+b+1)c+5c^2}}{2(a+b-c)} &
\textrm{otherwise (Id)}
\end{array}\right..\label{eq:u2}
\end{split}
\end{equation}
%\end{widetext}
All transitions are continuous but not differentiable between the
phases of I, II, and III, as well as between the four active
sub-phases. In the vicinity of all transition lines, either the
link density $\rho_\infty$ or the hole density $u_\infty$ vanishes
linearly ($\beta_1=\beta_2=1$).

It is interesting to note the emergence of a nontrivial phase
transition  between Ib and III in the region of $1-b<c<1+b$, where
the link creation and the link annihilation dynamics compete each
other. In this region, only the process representing the second
term of $W_+$ in Eq.~(\ref{tran_rate_k2}) is possible in the link
creation dynamics. Therefore, the link creation dynamics is
impossible when $\rho$ becomes zero (no-link state). Once the
system gets into the no-link state , it cannot escape out of that
state ({\em absorbing} state). Based on these observations, we can
map the dynamics onto the well-known {\em contact process}~\cite{cp}, which
exhibits an absorbing phase transition from vacuum (isolation
phase) into an active phase. This transition is known to belong to
the so-called {\em directed percolation} (DP) universality
class~\cite{abtr}.

Near the transition between Ib and III, we can easily write the
Langevin equation for small $\rho$ as
\begin{equation}\label{exact3}
\frac{d\rho}{dt}=3\varepsilon_1\rho -B\rho^2 +
\sqrt{\frac{4B\rho}{L}} \xi(t),
\end{equation}
where $\varepsilon_1=c_1^*-c$ with the transition point
$c_1^*=(1+2a-b)/3$ and $B=(1-a+2b)/3>0$. This equation is
identical to the MF Langevin equation describing the DP-type
absorbing phase transition, which is characterized by the noise
amplitude proportional to $\sqrt{\rho}$. The steady-state link density
behaves as $\rho_\infty\simeq (3/B)\varepsilon_1^{\beta_1}$ with
$\beta_1=1$. It is well known that the fluctuation exponent
$\gamma_1^\prime=0$ in the active side for the MF DP universality
class~\cite{abtr}. Utilizing the hyperscaling relation, we find
the FSS exponent $\bar\nu_1=2$, and thus expect $\rho_\infty\sim
L^{-\beta_1/{\bar\nu_1}}=L^{-1/2}$ at the transition~\cite{surv},
which is confirmed numerically (not shown here). The relaxation
time also diverges as $\tau\sim \varepsilon_1^{-\nu_\|}$ with
$\nu_\|=1$, in contrast to the case of the Ia-III transition line
where $\tau$ is finite.

The transition from Ib to II is also interesting. The
corresponding Langevin equation in the active side is given, in
terms of the hole density $u=1-\rho$, as
\begin{equation}\label{exact4}
\frac{du}{dt}=\varepsilon_2 u -3B u^2 + \sqrt{\frac{6B u^2}{L}}  \xi(t),
\end{equation}
where $\varepsilon_2=c-c_2^*$ with $c_2^*=a-b$. The steady-state
hole density scales as  $u_\infty\simeq
(1/(3B))\varepsilon_2^{\beta_2}$ with $\beta_2=1$. This type of
the multiplicative noise {\em linear} in the activity field $u$
has been studied extensively in literatures to describe various
physical systems including nonequilibrium wetting and
synchronization phenomena for spatially extended
systems~\cite{Munoz1998}. Naive power counting which is expected
to hold in the MF systems yields the FSS exponent $\bar\nu_2=1$
and $\gamma^\prime_2=-1$. Thus, in the limit of
$\varepsilon_2=0^+$, we have $u\sim L^{-1}$. As $u\sim L^{-1/2}$
in the paradise phase and at the transition ($\varepsilon_2\le
0$), we expect a discontinuous jump in the hole density for
finite systems at the Ib-II transition. Similar behaviors are
found near the transitions from Id and Ic to II. All these results
are confirmed numerically (not shown here).

Note that the exponent $\beta_2$ changes from
$1/k=1/2$ to $1$ with the interaction term ($b>0$). The relaxation
time exponent $\nu_{||}$ also changes from $1-1/k=1/2$ to $1$.
Inside the paradise phase II, the linear term (the first term in
the right-hand side of Eq.~(\ref{exact4})) vanishes and the hole
density decay exponent becomes $\delta=1$, which is the same as
$\delta=1/(k-1)=1$ at $b=0$. The FSS exponent $\bar\nu_2=1$ for
both $b=0$ and $b>0$.

%In fact, we find $\beta_2=1$ for any $k$ with $b>0$, which will be
%discussed later. We also find that the multiplicative factor of
%the noise is proportional to $\sqrt{u^k}$ for general $k$.

\subsubsection{$k\ge 3$}

%\begin{figure}[!t]
%\centerline{\includegraphics[scale=.8]{fig7_phasediagram_k3.eps}}
%\caption{(Color online) (a) The phase diagram for the case with
%$k=3$ and $b<\frac{1}{6}$, and (b) the mean field solution
%$u_3(a,c)$ when $b=0.1$. For the mixed phase II+III, we plot
%$1-\rho_{\textrm{uns}}$ rather than $0$ and $1$. The numerical
%simulations confirm our mean field expectation (not shown here).}
%\label{fig:rhosk3}
%\end{figure}

\begin{figure}[!t]
\centerline{\includegraphics[scale=.8]{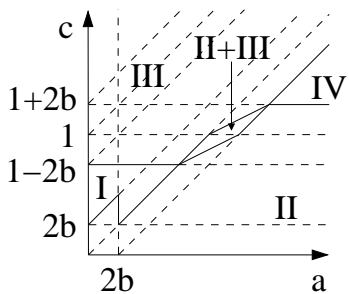}}
\caption{The phase diagram for the case with $k=3$ and
$0<b<\frac{1}{6}$. } \label{fig:rhosk3}
\end{figure}

For large $k$, we expect more complicated phase diagrams. For
example, see Fig.~\ref{fig:rhosk3} for the phase diagram at $k=3$
and small $b$. The parameter space is divided by 11 lines, such as
$c=1\pm 2b$ , $c=1$, $c=2b$, $c=a+(1\pm 2b)$, $c=a+1$, $c=a\pm
2b$, $c=a$, and $a=2b$. The last line of $a=2b$ is the boundary
line between regions where the $k$-hole constraint is strictly
valid and becomes loose. The divided regions are grouped into five
phases including one new phase ({\em inactive `mixed'} phase) in
the central region (Fig.~\ref{fig:rhosk3}).

In this mixed phase, the system reaches either the isolation phase or the paradise
phase, depending on the initial conditions and the
stochastic dynamics. This is similar to the absorbing phase
belonging to the {\em directed Ising} universality
class~\cite{Park} with two symmetric absorbing states. But, here,
there is no symmetry between absorbing states.

Our main interest lies in the active phase and the surrounding
inactive (isolation and paradise) phases. As expected, we find
that the active phase shrinks due to the expansion of both
inactive phases. But, the expansion of the paradise phase is both
qualitatively and quantitatively different in the regions of
$a>2b$ and $a<2b$  at $k=3$. For $a>2b$, the $k$-hole constraint
remains intact. The density decrease due to the $k$-hole
constraint happens to be balanced exactly by the density increase
due to the interactions at $k=3$, which results in maintaining the
I-II phase boundary $(c=a)$ as it is at $b=0$. For $k>3$, the
interaction term dominates and thus the paradise phase invades
into the active phase. For $a<2b$ where the $k$-hole constraint is
loose, the active phase shrinks more and the I-II phase boundary
is given by the $c=a+2b$ line.

The hole density in the phase II decays as $u(t)\sim t^{-\delta}$
with $\delta=1/2$ for $a>2b$ and $\delta=1$ for $a<2b$. For
general $k$, we find $\delta=1/[k/2]$ ($[x]$ is the integer value
of $x$) for $a<2b$. Extending to the other regions of
$2b(n-1)<a<2bn$ with an integer $n\le k-2$ is straightforward (not
reported here). For $a>2b(k-2)$ where the $k$-hole constraint is
valid, we find $\delta=1/(k-1)$ (the same as that at $b=0$).

Finally, we report the values of scaling exponents near the I-II
boundary. We find $\beta_2=1$, $\bar\nu_2=1$, $\nu_\|=2$ for
$a<2b$ and $\beta_2=1/2$, $\bar\nu_2=1/2$, $\nu_\|=1/2$ for $a>2b$.
Generalization to higher $k$ is straightforward. Near the I-III
boundary, we find $\beta_1=1$, which holds for any $k$.

\section{Bystander effects and social implications}\label{sect:bystander}

We investigate various bystander effects at the level of
society/community for general $k$, in relation to the empirical data
analysis by social psychologists. Our model is set up, based on the
experimental results in the laboratory, but the specific values of
control parameters cannot be directly inferred from these results.
We  compare our model study with corresponding field
studies~\cite{Amato1983,Levine1994}, in order to understand its
social implications properly.

\begin{figure*}[t]
\includegraphics[scale=1]{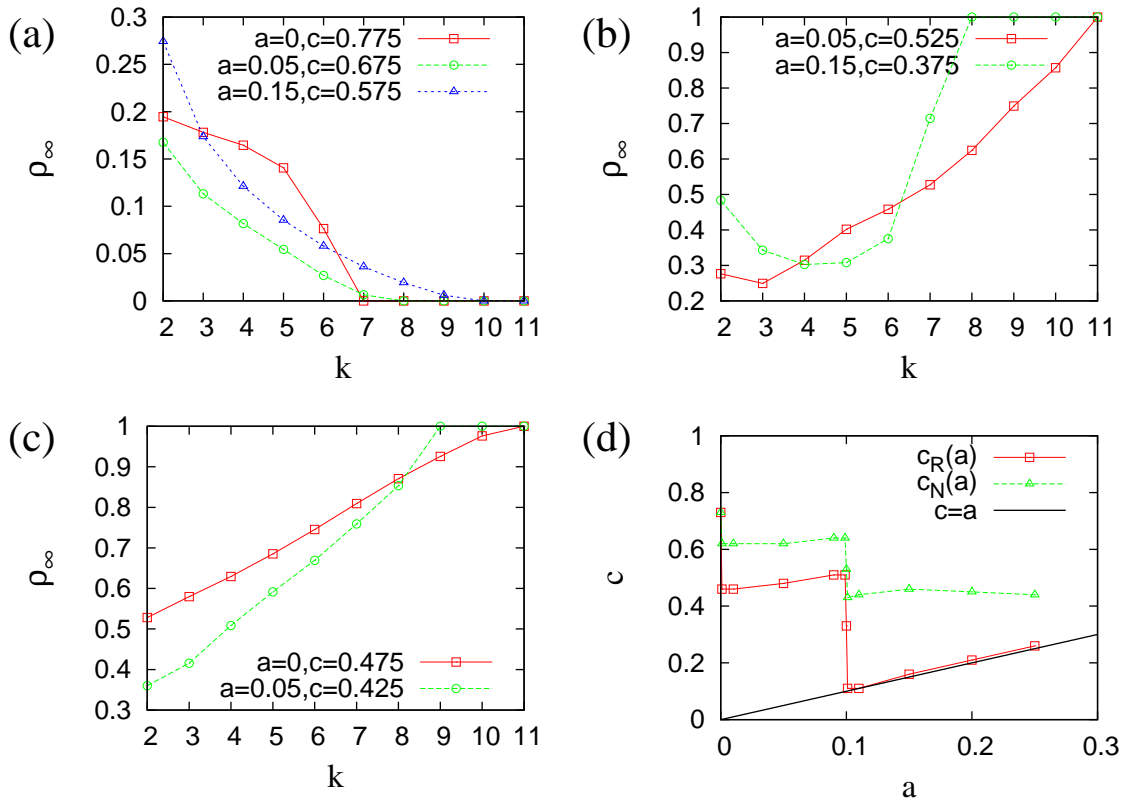}
\caption{(Color online) The various bystander effects: (a) NBE, (b)
CBE, and (c) RBE when $b=0.05$ and $a=0$, $b$, and $3b$. (d)
Critical potential cost lines $c_{\rm R}(a)$ and $c_{\rm N}(a)$ in
the $(a,c)$ phase diagram at fixed $b=0.05$. We find the NBE for
$c>c_{\rm N}(a)$ and the RBE for $c<c_{\rm R}$. In between these two
boundaries, the CBE is found. When identifying the BE's we consider
only the active phase, where only one stationary link density
exists.} \label{fig:rhokc}
\end{figure*}

The control parameters ($a,b,c,k$) can be regarded as given or
intrinsic to a society/community, hence depending on various factors
of its socioeconomic condition and also on the taxonomy of helping
behavior as well~\cite{Amato1983,Levine1994}. For the taxonomy of
helping, social psychologists introduced three independent
dimensions: doing (direct help) versus giving (indirect help),
spontaneous versus planned, and serious versus nonserious.
\textit{Helping measures}, such as helping rate, were observed from
field studies performed in the streets of $55$ cities and towns in
Australia~\cite{Amato1983} and of $36$ cities in
U.S.~\cite{Levine1994}.

For example, in one of experiments the experimenter walking on a
street drops a pen accidentally in front of a solitary pedestrian,
and continues walking, then sees whether the pedestrian picks the
pen and bring it to the experimenter or simply ignores. The
helping rate may correspond to the success rate in our model,
which is defined as the number of successful interventions divided
by the total number of interventions. In general the success rate
is a monotonic function of the link density $\rho$, i.e. the
aggregate expectation of help in a community. Therefore we can
interpret $\rho$ as a helping measure of a community. The helping
behavior in this experiment belongs to the direct, spontaneous,
and nonserious type with a low potential cost (small $c$), while
the rescue attempts in emergency situations are classified into
the direct, spontaneous, and serious type with large $c$.

The socioeconomic factors (\textit{community variables}) include
the population size, population density, cost of living,
unemployment rates, and etc. Here the number of witnesses $k$  per
accident corresponds to the population density (crowdedness) and
the system size $N$ to the population size, respectively. While
the population density and the population size are highly
correlated in reality (big cities are usually crowded), we
considered them as being independent to each other for generality.

The previous empirical studies showed that the most relevant
factors for the helping rate are the population density and the
population size. The strong negative correlation between the
helping rate and the population density/size was found. This
implies the normal bystander effect, i.e. $\rho$ decreasing with
$k$ in our model. However, in some nonserious experiments, the
helping rate first decreases and then increases as the population
size (or density) increases~\cite{Amato1983}, and a slightly
positive correlation between the helping rate and the population
density was also found~\cite{Levine1994}. These unusual bystander
effects can be supported by various behaviors of $\rho$ observed
in our model study (see below).

The experiments in the field studies have been done for measuring
the aggregate expectation of help in a given society, not for
simulating our model dynamics. We assume that the helping network
in the society has evolved by our model dynamics and has become
already stationary at the time of the experiments. By comparing
the experimental data with the stationary results of our model,
one may check the validity of our coevolving helping network (CHN)
model and also predict some social features in real social
systems. Especially, we focus on the various bystander effects
found in the field studies.

%Before going further we briefly note that the correspondence
%between our model study and field studies by social psychologists
%is not straightforward. One may say that the experimenter is
%always the stranger to subjects in the streets so that the value
%of $a$ in our model should be always zero. And the authors in the
%previous works did not show how many witnesses were involved in
%their experiments. Hence we need alternative point of view. Let us
%assume the helping network in a community has been evolved and
%becomes stationary in the same way as our model. Then, we measure
%the aggregate expectation of help of a community by observing the
%response of subjects to the experimenter. Only from the
%observables, the parameters $a,b,$ and $c$ can be inferred and
%then be interpreted properly.

Now we discuss the  $k$-dependent behavior of the link density
$\rho$, based on the numerical results for the CHN model, shown in
Fig.~\ref{fig:rhokc}. The behaviors can be categorized into three:
(i) the \textit{normal} bystander effect (NBE), i.e. $\rho$
decreasing with $k$, (ii) the \textit{reverse} bystander effect
(RBE), i.e. $\rho$ increasing with $k$, and (iii) the
\textit{complex} bystander effect (CBE), i.e. $\rho$ decreasing and
then increasing with $k$. With $b$ fixed, for each value of $a$,
there appear two critical values of the potential cost $c$,
separating the three different BE's. We observe the NBE phase for
$c>c_{\rm N}(a)$, the CBE phase for $c_{\rm R}(a)<c<c_{\rm N}(a)$,
and the RBE phase for $c<c_{\rm R}(a)$, respectively.

%When $a=0$, $c_{\rm P}(0)=c_{\rm N}(0)\equiv c_0$. And the values
%of $c_{\rm P}(a)$ and $c_{\rm N}(a)$ drastically change at the
%line $a=2b$.

%For $0<a\leq 2b$ we can define another critical potential cost
%$c_1'(a)$ by excluding the results for $k=2<k_c$.
%We can define $k_{\rm min}$ minimizing $\rho$ for given $c$. We
%also observe $k_{\rm min}$ increasing with $c$.

To understand how the various BE's appear as the potential cost
$c$ varies, we first consider the simplest case of $a=0$ in
Eq.~(\ref{eq:xvi}). For large $c$, the possibility of successful
intervention becomes small. This negative effect in the link
density (small $\rho$) is accelerated with $k$, because the
$k$-dependent interaction term becomes negative. Therefore, we
expect that $\rho$ decreases with $k$ (NBE) in the stationary
state. For small $c$, the tendency toward successful intervention
is accelerated with $k$ in a positive way. So we expect the RBE.

To be more specific, the dynamics can be simplified by the mean
field (MF) approximation of $x_{vi}$ at $a=0$:
\begin{equation}
\rho(t+\Delta t)=1+b(k-1)[2\rho(t)-1]-c,
\end{equation}
then the stationarity condition yields
\begin{equation}
\rho_\infty=\frac{1}{2}+\frac{c_0-c}{1-2b(k-1)} \label{eq:mf_xvi}
\end{equation}
with $c_0=1/2$. One can easily see the NBE phase for $c>c_0$, and
the RBE phase for $c<c_0$. Numerical simulation results are
consistent with this MF picture, except that $c_0\simeq 0.73$.

The positive $a$ enhances the possibility of selecting the witness
$i$ connected to the victim $(\rho_{iv}=1)$ as an intervener, and
hence the $k$-hole constraint comes into play. The $k$-hole
constraint, in general, suppresses the link density $\rho$. So we
expect that the NBE phase is not affected, because $a$ only
accelerates the decay of $\rho$ with $k$ faster.

However, the RBE phase may be modified significantly in the region
where the $k$-hole constraint becomes important. Here, the
interaction term proportional to $b(k-1)$ competes with $a$. For
sufficiently large $k$, we may ignore $a$ compared to the
interaction term, and then $\rho$ increases with $k$. However, for
small $k$, the effect of positive $a$ becomes strong so the $k$-hole
constraint may make $\rho$ to decrease with $k$ up to some value of
$k$. One may expect that this non-monotonic behavior of $\rho$ with
$k$ (CBE) appears between the NBE and the RBE phase
(Fig.~\ref{fig:rhokc}(d)).

Note that there is a big jump in the phase boundary at $a=2b$. As
discussed in the previous section, the $k$-hole constraint becomes
loose for $a<2b$, and becomes strong for $a>2b$. As a result, the
link density decreases discontinuously as one crosses the $a=2b$
point (see, e.g.,~Fig.~\ref{fig:rhosk3}). So it is easily expected
that the NBE phase expands and the RBE phase shrinks. Moreover, the
RBE phase effectively disappears for $a>2b$, because it is almost
squeezed down to the $c=a$ line where the paradise phase ($\rho=1$)
starts to appear.

The complex and reverse bystander effects found in our model can
support the unusual bystander effects found for the nonserious
experiments (small $c$). Moreover, when the relation between
witnesses is more important than the witness-victim relation ($a$ is
smaller than $b$), our model predicts that the CBE and the RBE are
more commonly observed in nonserious experiments. To confirm the
robustness of CBE and RBE in reality, more empirical/experimental
studies are necessary.

\section{\label{sect:con}Summary}

We have studied how the collective helping behavior and bystander
effects at the level of society emerge from the repeated
helper-recipient interactions by means of the coevolving helping
network model. Its dynamic rules are based on the results of
social psychological experiments. By analyzing the dynamics of the
network link density (aggregate expectation of help) and its
steady states and fluctuations, we present the full phase diagram
with various active and inactive phases and explore the nature of
phase transitions between them. Close to the transition between
the active and paradise phases, the critical behavior turns out to
depend on the number of witnesses per accident $k$. This is due to
the link creation suppressed by the $k$-hole constraint, which
also governs the algebraic decaying behavior in the paradise
phase.

We have numerically found various kinds (normal, reverse, and
complex) of collective bystander effects for large $k$ and proposed
the underlying mechanism. The normal and complex (non-monotonic)
bystander effects are consistent with the field study results for
the cases with serious and nonserious helping situations,
respectively. In addition, we expect the reverse bystander effect to
occur when the witness-victim relation is sufficiently less
important than the relation among witnesses for nonserious helping
situations.

\begin{acknowledgments}
HJ thanks Eunyoung Moon for helpful discussions.
\end{acknowledgments}

%\appendix
%\section{Appendixes}

%\newpage %Just because of unusual number of tables stacked at end
%\bibliography{apssamp}% Produces the bibliography via BibTeX.

\begin{thebibliography}{00}
\bibitem{exception}
There exists the exceptional region where $\rho_\infty$ for $b>0$
is larger than $\rho_\infty$ for $b=0$, which eventually
disappears as soon as $b\geq \frac{1}{4}$. This region is
surrounded by two lines and one curve:
$c=\frac{3}{4}a+\frac{1}{4}$,
$c=\left(\frac{1+2b+\sqrt{1-4b}}{2}\right)a
+\frac{1-2b-\sqrt{1-4b}}{2}$, and $a=\frac{2bc+b-2c^2+3c-1+
\sqrt{(c-1-b)(c-1+3b)}}{1+3b-c+\sqrt{(c-1-b)(c-1+3b)}}$.
\bibitem{surv}
In finite systems with absorbing states, the system will be always
trapped into one of absorbing states in the long-time limit. Thus,
it is well known that the average over surviving samples is a
useful quantity for the finite-size properties. Here, the order
parameter $\rho$ is averaged over surviving samples only.
\end{thebibliography}

\end{document}